\documentclass[12pt]{article}
\usepackage{subeqn}
\usepackage{epsfig}
\newcommand{\be}{\begin{equation}}
\newcommand{\ee}{\end{equation}}

\newcommand{\no}{\noindent}
\newcommand{\ce}{\begin{center}}
\newcommand{\nc}{\end{center}}

\makeatletter
\@addtoreset{equation}{section}
\makeatother

\def\sqr#1#2{{\vcenter{\vbox{\hrule height.#2pt
 \hbox{\vrule width.#2pt height#1pt \kern#1pt
 \vrule width.#2pt} \hrule height.#2pt}}}}

\def\operp{\hbox{${\kern+.25em{\bigcirc}
\kern-.85em\bot\kern+.85em\kern-.25em}$}}

\def\lsim{\;\raise0.3ex\hbox{$<$\kern-0.75em\raise-1.1ex\hbox{$\sim$}}\;}
\def\gsim{\;\raise0.3ex\hbox{$>$\kern-0.75em\raise-1.1ex\hbox{$\sim$}}\;}
\def\no{\noindent}

\def\ce{\centerline}
\def\ve{\vfill\eject}
\def\rdots{\mathinner{\mkern1mu\raise1pt\vbox{\kern7pt\hbox{.}}\mkern2mu
 \raise4pt\hbox{.}\mkern2mu\raise7pt\hbox{.}\mkern1mu}}

\def\e e{$e^+ e^-$ }

\begin{document}

\ce{\bf THE ELEMENTARY PARTICLES AS QUANTUM KNOTS}
\ce{\bf IN ELECTROWEAK THEORY}

\vskip.3cm

\ce{\it Robert J. Finkelstein}
\vskip.3cm

\ce{Department of Physics and Astronomy}
\ce{University of California, Los Angeles, CA 90095-1547}

\vskip1.0cm

\no {\bf Abstract}.  We explore a knot model of the elementary particles that
is compatible with electroweak physics.  The knots are quantized and their
kinematic states are labelled by $D^j_{mm^\prime}$, irreducible representations
of $SU_q(2)$, where $j=N/2,~m=w/2,~m^\prime=(r+1)/2$ and $(N,w,r)$ designate
respectively the number of crossings, the writhe, and the rotation of the knot.
The knot quantum numbers $(N,w,r)$ are related to the standard isotopic spin
quantum numbers $(t,t_3,t_0)$ by $(t=N/6,~t_3=-w/6,~t_0=-(r+1)/6)$, where
$t_0$ is the hypercharge.  In this model the elementary fermions are low lying
states of the quantum trefoil $(N=3)$ and the gauge bosons are ditrefoils
$(N=6)$.  The fermionic knots interact by the emission and absorption of bosonic
knots.  In this framework we have explored a slightly modified standard 
electroweak Lagrangian with a slightly modified gauge group which agrees closely but not entirely with standard electroweak theory.
\vskip4.0cm

\no UCLA/07/TEP/13

\ve

\section{The Knot Model.}

The possibility that the elementary particles are knots has been
suggested by many authors, going back as far as Kelvin.  There are many different field
theoretic ways of constructing classical knots.  In particular, a
model related to the Skyrme soliton has been described by Fadeev,
and by Fadeev and Niemi.$^1$  
Here we study a knot model that is independent of its particular field
theoretic realization.  In this model the field quanta of
standard electroweak theory are to be understood not as point
particles but as quantum knots defined by $SU_q(2)$.
Since any knot model must at least 
be compatible with electroweak physics and
must also be defined by its symmetry group namely
$SU_q(2)$, these are the minimal requirements of a knot model.  
Then the field quanta
are quantum knots; and the most elementary particles (the elementary fermions)
are the simplest quantum knots (the quantum trefoils).  There are indeed
four trefoils and there are four classes of elementary fermions with three
fermions in each class.  It is then possible to posit four fermionic quantum
trefoils, where each trefoil has three states of excitation.  Then the
electron, muon, and tauon appear as three states of excitation of the leptonic
trefoil.  In this model the fermionic knots interact by the emission and
absorption of bosonic knots.

\vskip.5cm

\section{The Characterization of Oriented Knots.}

Three-dimensional knots are described in terms of their projections onto a
two-dimensional plane where they appear as two-dimensional curves with 4-valent
vertices.  At each vertex (crossing) there is an overline and an underline.  We shall be interested here in oriented knots.
The crossing sign of the vertex is +1 or -1 depending on whether the
orientation of the overline is carried into the orientation of the underline
by a counter-clockwise or clockwise rotation respectively.  The sum of all the
crossing signs is termed the {\bf writhe}, $w$, a topological invariant.  There is a second topological invariant, the {\bf rotation}, $r$, the number
of rotations of the tangent in going once around the knot.

Let $K$ and $K^\prime$ be oriented knot diagrams with the same 
writhe and rotation

\ve
 
\begin{eqnarray*}
w(K) &=& w(K^\prime) \\
r(K) &=& r(K^\prime)
\end{eqnarray*}
Then $K$ is topologically equivalent (regularly isotopic) to 
$K^\prime$.

We may label an oriented knot by the number of crossings $(N)$,
its writhe $(w)$, and rotation $(r)$.  The writhe and rotation are
integers of opposite parity.

The symmetry algebra of the unoriented knot is $SL_q(2)$.  We shall
describe the oriented knot $(N,w,r)$ by the subgroup $SU_q(2)$.  To make its
connection with $SU_q(2)$ explicit, we may label a knot by the
elements of the irreducible representation of $SU_q(2)$ as follows:
\[
D^j_{mm^\prime} = D^{N/2}_{\frac{w}{2}\frac{r+1}{2}}
\]

\vskip.5cm

\section{The Quantum Mechanical Knot.}

A {\bf physical} knot as a classical dynamical system will have two topological
integrals of the motion: the writhe $(w)$ and the rotation $(r)$.
We shall consider systems where $N$, the number of crossings, is a
dynamical integral of the motion as well.
We shall assume that the quantum mechanical knot has the same
integrals of the motion as the classical knot.  Then we label the
states of the quantum mechanical knot by these same integrals of the
motion.  

Since the knot symmetry may be
represented by $SU_q(2)$, we may take the 
kinematical quantum states to be elements of the irreducible representations of
$SU_q(2)$.  These are designated by $D^j_{mm^\prime}$ where
$2j+1$ is the dimensionality.

To label the states $D^j_{mm^\prime}$ by the integrals of the motion
$(N,w,r)$ we set
\begin{eqnarray*}
j &=& \frac{N}{2} \\
m &=& \frac{w}{2} \\
m^\prime &=& \frac{r+1}{2}
\end{eqnarray*}

\no These linear relations between $(jmm^\prime)$ and $(N,w,r)$
are the simplest that permit half-integer representations and also
respect the difference in parity between $w$ and $r$.  Then
\[
D^j_{mm^\prime} = D^{N/2}_{\frac{w}{2}\frac{r+1}{2}}
\]
These are by definition the kinematical states of the quantum
mechanical knot $(N,w,r)$.

The procedure that we have just followed resembles that followed for a
quantum mechanical top.  There the integrals of motion are the
components of the angular momentum, the symmetry of the spherical 
spinning top is described by $SU(2)$, 
and the quantum mechanical states
are the irreducible representations of $SU(2)$, again labelled
$D^j_{mm^\prime}$, where the indices in that case refer to the
angular momentum.  For example, the quantum states of the 
``spinning electron" are labelled by the fundamental 
representation of $SU(2)$.

The quantum mechanical description of the knot is not complete at this
point however since there is as yet no Hamiltonian and there are no operators
for $(N,w,r)$, but these will be supplied in due course.

\vskip.5cm

\section{The Knot Algebra.}

One way of seeing that $SL_q(2)$ is the appropriate algebra of the
knot is to observe on the one hand 
that the Kauffman algorithm$^2$ (for generating the Kauffman or the
Jones polynomial that characterizes a knot) may be expressed
in terms of the matrix
\be
\epsilon_q = \left(
\begin{array}{cc}
0 & q^{-1/2} \\
-q^{1/2} & 0 
\end{array} \right) \qquad
\epsilon_q^2 = -1
\ee
and on the other hand that $\epsilon_q$ is also the invariant matrix of $SL_q(2)$ since
\be
T^t\epsilon_q T = T\epsilon_q T^t = \epsilon_q
\ee
where $T$ belongs to a two-dimensional representation of
$SL_q(2)$. 

We shall now describe this algebra.  Let 
\be
T = \left(
\begin{array}{cc}
a & b \\ c & d
\end{array} \right)
\ee

Then the matrix elements of $T$ satisfy the following algebra
\begin{center}
\begin{tabular}{llll}
$ab = qba$ \quad & $bd = qdb$ \quad & $ad-qbc = 1$ \quad & $bc = cb$ \\
$ac = qca$ \quad & $cd = qdc$ \quad & $da-q_1cb = 1$ \quad & $q_1 = q^{-1}$   \hskip2.5cm (A)
\end{tabular}
\end{center}

In the discussion of electroweak we need only the unitary
subalgebra obtained by setting
\begin{eqnarray*}
d &=& \bar a \\
c &=& -q_1\bar b
\end{eqnarray*}
Then $(A)$ reduces to the following
\begin{center}
\begin{tabular}{lll}
$ab = qba$ \qquad & $a\bar a + b\bar b = 1$ \qquad & $b\bar b = 
\bar bb$ \\
$a\bar b = q\bar ba$ \qquad & $\bar aa + q_1^2\bar bb = 1$ \qquad &
\hskip4.5cm $(A)^\prime$
\end{tabular}
\end{center}
For the physical applications we need the higher representations of
$SU_q(2)$.  The $2j+1$-dimensional unitary irreducible
representations of the $SU_q(2)$ algebra $(A)^\prime$ are
\be
D^j_{mm^\prime} = \sum_{s,t} A^j_{mm^\prime}(s,t)
\delta(s+t,n_+^\prime)a^sb^{n_+-s}\bar b^t\bar a^{n_--t}
\ee
where
\[
A^j_{mm^\prime}(s,t) = \left[{\langle n^\prime_+\rangle_1!~
\langle n^\prime_-\rangle_1!\over
\langle n_+\rangle_1!~\langle n_-\rangle_1!}\right] 
\left\langle\matrix{n_+\cr s}\right\rangle_1~
\left\langle\matrix{n_-\cr t}\right\rangle_1~q^{t(n_+-s+1)}
(-)^t 
\]
and
\[
\begin{array}{rcl}
n_\pm &=& j\pm m \\ n^\prime_\pm &=& j\pm m^\prime \\
\end{array} \quad
\left\langle\matrix{n \cr s}\right\rangle_1 =
{\langle n\rangle_1!\over \langle s\rangle_1!\langle n-s\rangle_1!}
\quad \langle n\rangle_1 = {q_1^{2n}-1\over q_1^2-1} 
\]

Every term of (4.4) contains a product of non-commuting factors
that may be reduced (after dropping numerical factors) to the form
\[
a^{n_a}\bar a^{n_{\bar a}}b^{n_b}\bar b^{n_{\bar b}}
\]
The $\delta$-function in (4.4) requires that
\begin{eqnarray}
n_a-n_{\bar a} &=& m+m^\prime \\
n_b-n_{\bar b} &=& m-m^\prime
\end{eqnarray}
where $n_a,n_b,n_{\bar a},n_{\bar b}$ are the exponents of
$a,b,\bar a,\bar b$, respectively.

These relations (4.5) and (4.6) hold for every term of (4.4) and
are independent of $j$.

\vskip.5cm

\section{Gauge Group $U_a(1)\times U_b(1)$ of the $SU_q(2)$ 
Algebra.$^4$}

The $SU_q(2)$ algebra $(A)^\prime$ is invariant under the following gauge
transformations
\begin{eqnarray}
& &a^\prime = e^{i\varphi_a}a \hskip2.0cm 
b^\prime = e^{i\varphi_b}b  \\
& &\bar a^\prime = e^{-i\varphi_a}\bar a \hskip1.7cm
\bar b^\prime = e^{-i\varphi_b}\bar b \nonumber
\end{eqnarray}
Then every term $\sim a^{n_a}\bar a^{n_{\bar a}}b^{n_b}\bar b^{n_{\bar b}}$ of $D^j_{mm^\prime}$ is multiplied by
\be
e^{i\varphi_a(n_a-n_{\bar a})}e^{i\varphi_b(n_b-n_{\bar b})}
= e^{i\varphi_a(m+m^\prime)}e^{i\varphi_b(m-m^\prime)}
\ee
by (4.5) and (4.6). Then (5.1) induces the following gauge transformations on
$D^j_{mm^\prime}$
\begin{eqnarray}
D^{j~~\prime}_{mm^\prime} &=& e^{i\varphi_a(m+m^\prime)}
e^{i\varphi_b(m-m^\prime)}D^j_{mm^\prime} \\
\bar D^{j~~\prime}_{mm^\prime} &=& e^{-i\varphi_a(m+m^\prime)}
e^{-i\varphi_b(m-m^\prime)}\bar D^j_{mm^\prime}
\end{eqnarray}
By analogy with the electric charge we may define two ``knot 
charges" $Q_a$ and $Q_b$ determined by the writhe and rotation as
follows:
\begin{eqnarray}
Q_a &\equiv& -k(m+m^\prime) = -k\left(\frac{w+r+1}{2}\right) \\
Q_b &\equiv& -k(m-m^\prime) = -k\left(\frac{w-r-1}{2}\right) 
\end{eqnarray}
then (5.3) and (5.4) become
\begin{eqnarray}
D^{j~~\prime}_{mm^\prime} &=& U_aU_bD^j_{mm^\prime} \\
\bar D^{j~~\prime}_{mm^\prime} &=& U_a^\star U_b^\star
\bar D^j_{mm^\prime}
\end{eqnarray}
where
\begin{eqnarray}
U_a &=& e^{-ik^{-1}Q_a\varphi_a} \\
U_b &=& e^{-ik^{-1}Q_b\varphi_b}
\end{eqnarray}
Then $U_a$ and $U_b$ are two independent gauge transformations on
the irreducible representations $D^j_{mm^\prime}$ of $SU_q(2)$
and therefore also on the quantum states 
$D^{N/2}_{\frac{w}{2}\frac{r+1}{2}}$ of the knot $(N,w,r)$.  By 
(5.8) $D^j_{mm^\prime}$ and $\bar D^j_{mm^\prime}$ have opposite
charges.

We shall now examine the possibility that the elementary particles
are quantum knots described by the kinematic states
$D^{N/2}_{\frac{w}{2}\frac{r+1}{2}}$.

\section {The Knot Conjecture.}

We shall explore the possibility that
the elementary particles are quantum knots (knotted flux tubes).
In such a model the simplest knots (trefoils) should correspond to 
the most elementary particles (fermions).  Indeed
there are 4 trefoils and there are 4 families of fermions.
Each trefoil is labelled by $(N,w,r)$ where $N=3$ and each family of fermions is labelled by $(t,t_3,Q)$ where $t = 1/2$.
Is there a meaningful correspondence $(N,w,r) \leftrightarrow
(t,t_3,Q)$? i.e. 
\begin{description}
\item[(a)] between single trefoil solitons and single fermion
families?
\item[(b)] between states of a trefoil soliton and members of
a fermion family?
\end{description}

We may try to establish this correspondence by labelling both the
trefoil solitons and the fermion families by the irreducible
representations of $SU_q(2)$, namely $D^j_{mm^\prime}$ as follows:
\begin{eqnarray*}
& &(Nwr)\quad \leftrightarrow \quad jmm^\prime \quad \leftrightarrow
\quad~~ (t,t_3,Q) \\
& &{} \quad \uparrow \qquad ~~~\qquad \uparrow \qquad\qquad~~\quad 
\uparrow \\
& &{\rm trefoils}\quad\leftrightarrow ~~ SU_q(2)~~ \leftrightarrow ~~
\mbox{fermion family}
\end{eqnarray*}
where $D^j_{mm^\prime} = D^{N/2}_{\frac{w}{2}\frac{r+1}{2}}$ and
$N=3, j=\frac{3}{2}$, and $t = \frac{1}{2}$.

\vskip.5cm

\no {\it The Fermionic Solitons as Trefoils.}$^2$
\vskip.3cm

In order to put each trefoil in correspondence with one class of
fermions we shall compare the knot charges, $Q_a = -k(m+m^\prime)$, of the 4 trefoils with the electric charges of the 4 classes as shown
in Table 1:
\vskip.3cm

\ce{\bf Table 1.}
\vskip.2cm

\begin{center}
\begin{tabular}{ccc||cc}
\underline{\mbox{Trefoils}~~$(w,r)$} & 
\underline{$D^{N/2}_{\frac{w}{2}\frac{r+1}{2}}$}
& \underline{$Q_a$} & \underline{Fermion Class} &
\underline{$Q_f$} \\
(-3,2) & $D^{3/2}_{-\frac{3}{2}\frac{3}{2}}$ & 0 &
$(\nu_e\nu_\mu\nu_\tau)$ & 0 \\
(3,2) & $D^{3/2}_{\frac{3}{2}\frac{3}{2}}$ & $-3k$ &
$(e^-,\mu^-,\tau^-)$ & $-e$ \\
(3,-2) & $D^{3/2}_{\frac{3}{2}-\frac{1}{2}}$ & $-k$ &
$(d,s,b)$ & $-\frac{1}{3}~e$ \\
(-3,-2) & $D^{3/2}_{-\frac{3}{2}-\frac{1}{2}}$ & $2k$ &
$(u,c,t)$ & $\frac{2}{3}~e$ 
\end{tabular}
\end{center}
\vskip.3cm

where $N=3$ and
\begin{eqnarray}
Q_a &\equiv& -k(m+m^\prime) \\
Q_f &=& \mbox{electric charge of the fermion class}
\end{eqnarray}

There is a unique mapping and
single value of $k$ that permits one to match the
trefoils with the fermion classes by satisfying
\be
Q_a(w,r) = Q_f~\mbox{(fermion class)}
\ee
namely
\be
k = \frac{e}{3}
\ee
Then
\be
Q_a = -\frac{e}{3}~(m+m^\prime)
\ee
or
\be
Q_a = -\frac{e}{6}~(w+r+1)
\ee
Then $Q_a$ may be considered the electric charge of the quantum 
trefoil $D^{3/2}_{\frac{w}{2}\frac{r+1}{2}}$.

It is interesting that this same correspondence between trefoils and 
fermion classes was found in our earlier phenomenological work$^{3,4}$ before
the natural appearance of $Q_a$ as $-\frac{e}{3}~(m+m^\prime)$ or
$\left(-\frac{e}{6}~(w+r+1)\right)$ was noticed.
Other mappings of the trefoils onto the fermion families are possible,
but there is only one mapping with a single value of $k$.

One therefore identifies $Q_a$ with the electric charge.  One could
also attempt the match with $Q_b$, but in that case the neutrinos
would be assigned to the (3,2) knot, in contradiction to earlier
work$^{3,4}$ that associates neutrinos with (-3,2).  We defer the
interpretation of $Q_b$.

Since $Q_a \sim m+m^\prime = n_a-n_{\bar a}$, note that the vanishing
of $Q_a$ implies
\[
n_a = n_{\bar a}
\]
and therefore that $a$ and $\bar a$ may be eliminated from every
term of $D^j_{mm^\prime}$ with the aid of
\be
a^n\bar a^n = \prod_{s=0}^{n-1} (1-q^{2s}b\bar b) \qquad n\geq 1
\ee
as follows from $(A)^\prime$.  Therefore electrically neutral states (neutrinos
and neutral bosons) lie entirely in the $(b,\bar b)$ subalgebra.

Note also that $\bar D^j_{mm^\prime}$ has opposite charges from
$D^j_{mm^\prime}$ and may therefore be identified as the state
of the antiparticle.

Given the match in Table 1 we may now compare 
all the quantum numbers
$(t,t_3,Q)$ labelling the different classes of fermions in the
standard representation with the quantum numbers $(N,w,r)$ labelling
the corresponding quantum knots.
\vskip.3cm

\ce{\bf Table 2.}

\vskip.2cm

\begin{center}
\begin{tabular}{lccc||cccc}
\multicolumn{4}{c}{Standard Representation} & \multicolumn{4}{c}
{Knot Representation} \\
& \underline{$t$} & \underline{$t_3$} & \underline{$Q$} & 
\underline{$w$} & \underline{$r$} & 
\underline{$D^{3/2}_{\frac{w}{2}\frac{r+1}{2}}$}
& \underline{$Q_a$} \\
$(e\mu\tau)_{\rm L}$ & $\frac{1}{2}$ & $-\frac{1}{2}$ & $-e$ & 
$3$ & $2$ & $D^{3/2}_{\frac{3}{2}\frac{3}{2}}$ & $-e$ \\
$(\nu_e\nu_\mu\nu_\tau)_{\rm L}$ & $\frac{1}{2}$ & $\frac{1}{2}$ & $0$ &
$-3$ & $2$ & $D^{3/2}_{-\frac{3}{2}\frac{3}{2}}$ & $0$ \\
$(dsb)_{\rm L}$ & $\frac{1}{2}$ & $-\frac{1}{2}$ & $-\frac{1}{3}~e$ &
$3$ & $-2$ & $D^{3/2}_{\frac{3}{2}-\frac{1}{2}}$ & $-\frac{1}{3}~e$ \\
$(uct)_{\rm L}$ & $\frac{1}{2}$ & $\frac{1}{2}$ & $\frac{2}{3}~e$ &
$-3$ & $-2$ & $D^{3/2}_{-\frac{3}{2}-\frac{1}{2}}$ & $\frac{2}{3}~e$
\end{tabular}
\end{center}

One then reads off the following relations from Table 2.
\be
t = \frac{N}{6}
\ee
since $N=3$ for trefoils.  

Also $t_3$ is proportional to $w$ (not to $r$) and
\be
t_3 = -\frac{w}{6}
\ee
Since $m = \frac{w}{2}$
\be
t_3 = -\frac{m}{3}
\ee

Finally in the knot representation the electric charge is
\be
Q = -\frac{e}{3}~(m+m^\prime)
\ee
But in the standard theory (point particle representation)
\be
Q = (t_3+t_0) e
\ee

Since (6.11) and (6.12) must agree, we have
\be
t_3 + t_0 = -\frac{1}{3}~(m+m^\prime)
\ee
By (6.10) and (6.13) the hypercharge is
\be
t_0 = -\frac{1}{3}~m^\prime
\ee
Therefore alternative forms of the quantum state of the fermionic
knots are
\be
D^{N/2}_{\frac{w}{2}\frac{r+1}{2}} \qquad \mbox{or} \qquad
D^{3t}_{-3t_3~-3t_0}
\ee
Therefore the invariance group of the algebra, namely, $U_a(1)\times
U_b(1)$ defines the charge and hypercharge.  By (5.8) the adjoint
representations carry opposite charge.

The relation that we have just found between isotopic spin and knot 
quantum numbers is
\begin{eqnarray}
t~ &=& \frac{N}{6} \nonumber \\
t_3 &=& -\frac{w}{6} \\
t_0 &=& -\frac{r+1}{6} \nonumber
\end{eqnarray}

Table 2 refers to the $L$-chiral field.  The $R$-chiral field is 
unknotted since $t=0$.

\vskip.5cm

\section{Quantum Operators for $(N,w,r)$ and $Q$.}

Denote the operators whose eigenvalues are $(N,w,r+1)$ by
$({\cal{N}},{\cal{W}},{\cal{R}})$, i.e.
\begin{eqnarray}
{\cal{N}}~D^{N/2}_{\frac{w}{2}\frac{r+1}{2}} &=& 
N~D^{N/2}_{\frac{w}{2}\frac{r+1}{2}}(a,\bar a,b,\bar b) \\
{\cal{W}}~D^{N/2}_{\frac{w}{2}\frac{r+1}{2}} &=&
w~D^{N/2}_{\frac{w}{2}\frac{r+1}{2}}(a,\bar a,b,\bar b) \\
{\cal{R}}~D^{N/2}_{\frac{w}{2}\frac{r+1}{2}} &=&
(r+1)~D^{N/2}_{\frac{w}{2}\frac{r+1}{2}}(a,\bar a,b,\bar b)
\end{eqnarray}
where the $D^{N/2}_{\frac{w}{2}}$ are functions of 
$(a,\bar a,b,\bar b)$.  These operators have the following forms in
the basic algebra:
\begin{eqnarray}
{\cal{N}} &=& \omega_a+\omega_{\bar a}+\omega_b+\omega_{\bar b}\\
{\cal{W}} &=& \omega_a-\omega_{\bar a}+\omega_b-\omega_{\bar b}\\
{\cal{R}} &=& \omega_a-\omega_{\bar a}-\omega_b+\omega_{\bar b}
\end{eqnarray}
where the $\omega_x$ are dilatation operators defined by their action on every
term of $D^j_{mm^\prime}$ according to
\be
\omega_x(\ldots x^{n_x}\ldots) = n_x(\ldots x^{n_x}\ldots) \qquad
x = (a,\bar a,b,\bar b)
\ee
i.e. $\omega_x$ acts like $x~\frac{\partial}{\partial x}$.

The four knot eigenstates of ${\cal{N}}$ with eigenvalue $N=3$ are the
four trefoils $D^{3/2}_{\frac{w}{2}\frac{r+1}{2}}$.

The charge operator is
\begin{eqnarray}
{\cal{Q}} &=& -\frac{e}{3} \frac{{\cal{W}}+{\cal{R}}}{2} \quad
\left(= -\frac{e}{3}~(m+m^\prime)\right) \\
&=& -\frac{e}{3}~(\omega_a-\omega_{\bar a})
\end{eqnarray}

Then
\begin{eqnarray}
{\cal{Q}}~D^{3t}_{-3t_3~-3t_0} &=& -\frac{e}{3}~
(\omega_a-\omega_{\bar a}) D^{3t}_{-3t_3~-3t_0} \\
&=& -e(t_3+t_0) D^{3t}_{-3t_3~-3t_0}
\end{eqnarray}
The 3 eigenfunctions of ${\cal{Q}}$ with $N=3$ and $t=\frac{1}{2}$ are
$D^{3/2}_{-3t_3~-3t_0}$ (the four trefoils) and the 4 eigenvalues
$-e(t_3+t_0)$ are the charges of the 4 solitons.

One may compare (7.10) and (7.11) with similar equations for the
angular momentum of a top, namely
\begin{eqnarray}
L_3~D^j_{mm^\prime} &=& \frac{\hbar}{i}\left(x\frac{\partial}
{\partial y} - y\frac{\partial}{\partial x}\right)
D^j_{mm^\prime} \\
&=& -m\hbar D^j_{mm^\prime} \\
K_3~D^j_{mm^\prime} &=& -m^\prime \hbar D^j_{mm^\prime}
\end{eqnarray}
According to (7.10) and (7.12) the charge is quantized in units of 
$\frac{e}{3}$ while the angular momentum is quantized in 
units of $\hbar$.

The complete $L$-chiral wave function of the fermionic soliton is now
\be
F(\vec p,\vec s,\vec t) D^{3t}_{-3t_3~-3t_0}
(q|a\bar ab\bar b)
\ee
where the first factor is the $L$-chiral
standard Dirac wave function for a
point particle with momentum $\vec p$, spin $\vec s$, and isotopic
spin $\vec t$ and where the second factor is the internal knot
state.

\vskip.5cm

\section{The State Space of the Solitons.}

The 4 solitons are eigenstates of ${\cal{N}}, {\cal{W}}$, and
${\cal{R}}$, but they are functions of $(a,\bar a,b,\bar b)$.  
They may be numerically evaluated on
the state space of the algebra.  Let us next describe this state
space.

Since $b$ and $\bar b$ commute, they have common eigenvalues.  Let
$|0\rangle$ be designated as a ground state and let
\begin{eqnarray}
b|0\rangle &=& \beta|0\rangle \\
\bar b|0\rangle &=& \beta^\star|0\rangle
\end{eqnarray}
and
\be
\bar bb|0\rangle = |\beta|^2|0\rangle
\ee
where $\bar bb$ is Hermitian with real eigenvalues and orthogonal
eigenstates.

One finds by $(A)^\prime$ that
\be
\bar bb|n\rangle = E_n|n\rangle
\ee
where
\be
|n\rangle \sim\bar a^n|0\rangle
\ee
and
\be
E_n = q^{2n}|\beta|^2
\ee
$\bar bb$ resembles the Hamiltonian of an oscillator but with 
eigenvalues arranged in geometrical progression and with $|\beta|^2$
corresponding to $\frac{1}{2}~\hbar\omega$.  If we take $H(\bar bb)$ 
to be the Hamiltonian of the knot, where the functional form of $H$
is left unspecified, it will have the same eigenstates as $\bar bb$.

Here $\bar a$ and $a$ are raising and lowering operators 
respectively
\begin{eqnarray}
\bar a|n\rangle &=& \lambda_n|n+1\rangle \\
a|n\rangle &=& \mu_n|n-1\rangle
\end{eqnarray}
One finds
\begin{eqnarray}
|\lambda_n|^2 &=& 1-q^{2n}|\beta|^2 \\
|\mu_n|^2 &=& 1-q^{2(n-1)}|\beta|^2
\end{eqnarray}
If there is a highest state $(M)$ then
\be
1-q^{2M}|\beta|^2 = 0
\ee
If there is a lowest state $(M^\prime)$ then
\be
1-q^{2(M^\prime-1)}|\beta|^2 = 0
\ee
If there are both a highest and lowest state, then
\be
q^{2M} = q^{2(M^\prime-1)}
\ee
and if $q$ is real as we assume
\be
M = M^\prime -1
\ee
but this is not possible since $M^\prime \leq M$.  Therefore if
$q$ is real, there is no finite representation of the elements of
this algebra.  This may be a lowest or a highest state but not
both.

For the physical application however, we may require finite
representations.  If this is so, it must be possible to cut off the
$q$-oscillator spectrum by imposing physical boundary conditions
at one or both of the upper and lower bounds.  Indeed, insofar as the present
model is an electroweak model that excludes gluon and gravitational forces,
one must expect that the neglected physics will impose boundary conditions on this model.

\vskip.5cm

\section{The Fermion States.}

We shall propose that the separate fermion states are the
ground and low lying excited states of the fermionic soliton and their
state functions are
\be
D^{3t}_{-3t_3~-3t_0}|n\rangle \qquad n = 0,1,2
\ee
where $|n\rangle$ is the $n^{\rm th}$ level of the $q$ oscillator.
If there are only 3 fermions in each family, we shall assume that
they occupy the lowest levels so that $n$ takes on the values 
0,1,2.  We interpret $|0\rangle$ as the state of lowest energy.

Then the complete $L$-chiral wave function of a fermion becomes, by (7.15) and (9.1)
\be
F(\vec p,\vec s,\vec t) D^{3t}_{-3t_3~-3t_0}|n\rangle
\ee
For the internal state of the antiparticle we propose the adjoint
representation, namely
\be
\bar D^{3/2}_{-3t_3~-3t_0}|n\rangle
\ee
By choosing the adjoint, we guarantee that the charge of the
antisoliton is opposite to that of the soliton as shown in (3.9).

The states $|n\rangle$ are not only the eigenstates of the 
$q$-oscillator but they are also eigenstates of any Hamiltonian of the
form
\[
H = f(\bar bb)
\]
This $H$ commutes with the integrals of the motion:
\[
[H,{\cal{N}}] = [H,{\cal{W}}] = [H,{\cal{R}}] = 0
\]
as required. 

The relative masses of the fermions are determined by an effective
Hamiltonian $H(\bar bb)$.  Let the energy of the state $\psi_n =
D^{3t}_{-3t_3~-3t_0}|n\rangle$ be $E_n$.  Then
\be
H(\bar bb)\psi_n = E_n\psi_n
\ee
where the $E_n$ determines the relative masses of the Fermions.
At the field-theoretic level the masses of the Fermions are
determined by their various interactions with other fields, but
within the limitations of the knot model we may assume that there is
an effective $H$ of the following unknown form:
\be
H = f(\bar bb)
\ee
Then
\begin{eqnarray}
H\psi_n &=& f(\bar bb) D^j_{mm^\prime}|n\rangle \nonumber \\
&=& D^j_{mm^\prime}f(q^{6Q/e}\bar bb)|n\rangle
\end{eqnarray}
where $Q$ is the electric charge of $D^j_{mm^\prime}$ and differs
among the four solitons.  Then
\begin{eqnarray}
H\psi_n &=& D^j_{mm^\prime}f(q^{6Q/e}q^{2n}|\beta|^2)|n\rangle 
\nonumber\\
&=& f(q^{6Q/e+2n}|\beta|^2)\psi_n
\end{eqnarray}
and
\be
E_n = f(q^{6Q/e}q^{2n}|\beta|^2)
\ee
By choosing $H$ to agree with the effective mass term in the
Higgs Lagrangian of the standard theory, one arrives at $^{3,4}$
\be
f(\bar bb) = \bar D^j_{mm^\prime} D^j_{mm^\prime}
\ee

\section{The Fermion-Boson Interactions.}

In this model the fermion solitons interact by the emission and absorption of
bosonic solitons.
We denote the generic fermion-boson interaction by
\be
\bar{\cal{F}}_3{\cal{B}}_2{\cal{F}}_1
\ee
where
\begin{eqnarray}
{\cal{F}}_1 &=& F_1(\vec p,\vec s,\vec t)D^{3/2}_{m_1m_1^\prime}
(a\bar ab\bar b)|n_1\rangle \\
\bar{\cal{F}}_3 &=& \langle n_3|\bar D^{3/2}_{m_3m_3^\prime}
(a\bar ab\bar b)\bar F_3(\vec p,\vec s,\vec t) \\
{\cal{B}}_2 &=& B_2(\vec p,\vec s,\vec t)D^j_{m_2m_2^\prime}
(a\bar ab\bar b)
\end{eqnarray}
Here $F(\vec p,\vec s,\vec t)$ and $B(\vec p,\vec s,\vec t)$ are
the standard fermionic and bosonic normal modes.  Then (10.1)
becomes
\be
(\bar F_3B_2F_1) \langle n_3|\bar D^{3/2}_{m_3m_3^\prime}
D^j_{m_2m_2^\prime}D^{3/2}_{m_1m_1^\prime}|n_1\rangle
\ee
The correction to the standard matrix elements appears in the
second factor, namely
\be
\langle n_3|\bar D^{3/2}_{m_3m_3^\prime}D^j_{m_2m_2^\prime}
D^{3/2}_{m_1m_1^\prime}|n_1\rangle
\ee
If there are $M$ generations of fermions, then $n_1$ and $n_3$ take on
values $0\ldots M-1$.

We must require that the basic internal interaction be invariant
under gauge transformations, $U_a(1)\times U_b(1)$, of the 
underlying algebra, i.e.
\be
(\bar D^{3/2}_{m_3m_3^\prime})^\prime~
(D^j_{m_2m_2^\prime})^\prime~
(D^{3/2}_{m_1m_1^\prime})^\prime =
\bar D^{3/2}_{m_3m_3^\prime}D^j_{m_2m_2^\prime}
D^{3/2}_{m_1m_1^\prime}
\ee
Here both charges are conserved:
\begin{eqnarray} 
exp[ik\varphi_a](-Q_a(3)+Q_a(2) + Q_a(1)) &=& 1 \\
exp[ik\varphi_b](-Q_b(3)+Q_b(2)+Q_b(1)) &=& 1
\end{eqnarray}
Therefore both charges are conserved:
\be
Q(1) + Q(2) = Q(3)
\ee
Then
\begin{eqnarray}
m_3 &=& m_1+m_2 \\
m_3^\prime &=& m_1^\prime + m_2^\prime
\end{eqnarray}
and the possible values of $(j,m,m^\prime)$ for the intermediate
boson are restricted by the known values of $(j,m,m^\prime)$ for
the initial and final fermions.

If the rules for connecting $m$ and $m^\prime$ to $t_3$ and $t_0$ are 
extended without change from fermions to the intermediate boson, then
the conservation of $Q_a$ and $Q_b$ by the basic interaction
implies the conservation of $t_3$ and $t_0$ by the same interaction.
Therefore we adopt for bosonic knots the same rules as for fermionic
knots:
\be
\begin{array}{ccc}
m = -3t_3 &  & w = -6t_3 \\
m^\prime = -3t_0 &  \mbox{or} & r+1 = -6t_0 \\
j = 3t &  &  N = 6t  
\end{array}
\ee

Applied to the vector bosons these rules imply Table 3:

\ce{\bf Table 3.}
\vskip.3cm

\[
\begin{array}{ccrcc}
& \underline{t} & \underline{t_3} & \underline{t_0} &
\underline{D^{3t}_{-3t_3~-3t_0}} \\
W^+ & 1 & 1 & 0 & D^3_{-3~0} \\
W^- & 1 & -1 & 0 & D^3_{3~0} \\
W^3 & 1 & 0 & 0 & D^3_{0~0} \\
W^0 & 0 & 0 & 0 & D^0_{0~0}
\end{array}
\]
The first 3 columns of the Table express the fact that $\vec W$ is
an isotriplet and $W^0$ is an isosinglet in the standard theory.

The fourth column $D^{3t}_{-3t_3~-3t_0}$ labels the internal 
states of the four vector bosons.  
If $t=1$, $j=3$; and if $j = \frac{N}{2}$, as we have assumed, 
then $N=6$
and $\vec W$ is a ditrefoil consistent with the pair production of
fermions by $\vec W$.$^3$

Since $W^0$ is coupled only to $U(1)$ in the standard theory, there
is no self-coupling, i.e., it itself carries neither electric nor
hypercharge.  The assignment of $j$ to $W^0$ is also restricted in
the internal matrix element by the $q$-Clebsch-Gordan rules.  If
$j=0$ and we maintain the relation $j=N/2$, then$N=0$ and
$W^0$ is an unknotted clockwise loop.

The possiblity of extending the conservation laws and the same rule for
associating $Q_a$ and $Q_b$ with $(m,m^\prime)$ to all solitons
depends on the fact that $Q_a$ and $Q_b$ are independent of $j$.

The conservation of $Q_a$ and $Q_b$, i.e. the invariance of the action
under $U_a(1)\otimes U_b(1)$ is more fundamental in the present model
than the conservation of $t_3$ and $t_0$.  Electric charge
and isospin in this model are simply characterizations of the geometry
of the knotted soliton.

The complete matrix elements are of the following form:
\[
(\bar F_3B_2F_1)\langle n_3|\bar D^{3/2}_{m_3m_3^\prime}
D^{j_2}_{m_2m_2^\prime}D^{3/2}_{m_1m_1^\prime}|n_1\rangle
\]
Corrections to the standard matrix elements appear in the second
factor and have been computed in the phenomenological work.$^4$
These corrections are small, however, and it is necessary to 
make more refined calculations and also to go to 
higher order to judge their significance.  In addition some of the assumptions made in the earlier work may be dropped and
some need to be modified in light of the present paper.

We have therefore explored a field theoretic formulation of these
ideas.  The following sections repeat much of Ref. 5 and are
included for completeness.

\vskip.5cm

\section{Field Theory.$^5$}

In passing from the standard to the knot field theory we shall
modify both the symmetry group and the Lagrangian.  The symmetry
of standard electroweak local is $SU(2)\otimes$
local $U(1)$.
We shall assume a slightly expanded symmetry to characterize the knot
model, namely: 
\[
[\mbox{local}~ SU(2)\otimes ~\mbox{local}~
U(1)] \otimes [\mbox{global}~ (U_a(1)\otimes U_b(1))]
\]
The vector connection in the standard theory is 
\be
W_+t_++W_-t_- + W_3t_3+W_0t_0
\ee
where $(t_+,t_-,t_3,t_0)$ are the generators of the standard 
electroweak theory in the charge representation and 
\be
t_+ = \left(
\begin{array}{cc}
0 & 1 \\ 0 & 0 
\end{array} \right),~t_- = \left(
\begin{array}{cc}
0 & 0 \\ 1 & 0 
\end{array} \right),~t_3 = \left(
\begin{array}{cc}
1 & 0 \\ 0 & -1 
\end{array} \right),~t_0 = \left(
\begin{array}{cc}
1 & 0 \\ 0 & 1 
\end{array} \right)
\ee

We now replace (11.1) by
\be
W_+\tau_+ + W_-\tau_- + W_3\tau_3 + W_0\tau_0
\ee
where
\be
\tau_k = c_k(q,\beta) t_k{\cal{D}}_k \qquad
k = (+,-,3,0)
\ee
and the ${\cal{D}}_k$ are the charge states of the four vector mesons
\begin{eqnarray*}
{\cal{D}}_+ &\equiv& D^3_{-30}/N_+ = \bar b^3\bar a^3 \\
{\cal{D}}_- &\equiv& D^2_{03}/N_- = a^3b^3 \\
{\cal{D}}_3 &\equiv& D^3_{00} = f_3(\bar bb) \\
{\cal{D}}_0 &\equiv& D^0_{00} = 1
\end{eqnarray*}

The $c_k$ are
numerical functions of the parameters $(q,\beta)$ that are fixed
by relations between the masses of the vector bosons.

Here are the two-dimensional representation of the new generators:
\be
(\vec\tau,\tau_0) = \left(
\begin{array}{cc}
0 & {\cal{D}}_+ \\ 0 & 0 
\end{array} \right),~\left(
\begin{array}{cc}
0 & 0 \\ {\cal{D}}_- & 0
\end{array} \right),~\left(
\begin{array}{cc}
{\cal{D}}_3 & 0 \\ 0 & -{\cal{D}}_3
\end{array} \right),~\left(
\begin{array}{cc}
{\cal{D}}_0 & 0 \\ 0 & {\cal{D}}_0
\end{array} \right)
\ee

For the fermions and Higgs-like fields one replaces the numerically
valued 2-rowed spinors of isotopic spin $SU(2)$ by the following
operator valued spinors
\be
\left(
\begin{array}{c}
D_\nu^{3/2} \\ 0
\end{array} \right)~,~\left(
\begin{array}{c}
0 \\ D_\ell^{3/2}
\end{array} \right)~,~\left(
\begin{array}{c}
D_u^{3/2} \\ 0
\end{array} \right)~,~\left(
\begin{array}{c}
0 \\ D_d^{3/2} 
\end{array} \right)
\ee
where $D_r^{3/2}$ is an abbreviation for the irreducible
representation associated with the $r^{\rm th}$ soliton and where
$r = (\nu,\ell,u,d)$.  

We also introduce $\psi_{Ari}$ defined by
\begin{eqnarray}
\psi_{1ri} &=& \psi_{1r}~D_r(a\bar ab\bar b)|i\rangle~~
A=1~~r = \nu,\ell~~i=1,2,3 \\
\psi_{2ri} &=& \psi_{2r}~D_r(a\bar ab\bar b)|i\rangle~~
A=2~~r=u,d~~i=1,2,3
\end{eqnarray}
where the lepton and neutrino solitons are combined into one isotopic
spinor, $\psi_{1ri}$, and the up and down quarks into a second
isotopic spinor, $\psi_{2ri}$ and where $i$ runs over the three
states of the soliton.  Then
\be
\psi_1 = \left(
\begin{array}{c}
\psi_\nu~D_\nu^{3/2}|i\rangle \\ \psi_\ell~D_\ell
^{3/2}|i\rangle
\end{array} \right) \quad \mbox{and} \quad \psi_2 = \left(
\begin{array}{c}
\psi_u~D_u^{3/2}|i\rangle \\ \psi_d~D_d^{3/2}|i\rangle
\end{array} \right)
\ee

\vskip.5cm

\section{$\tau$ -Commutators, Gauge Fields, Field Strengths and
Interactions.}
\no (a) {\bf $\tau$-Commutators.}

\be
\tau_k = c_kt_k{\cal{D}}_k \qquad k = (+,-,3)
\ee
\begin{eqnarray}
[\tau_k,\tau_\ell] &=& c_kc_\ell[t_k{\cal{D}}_k,t_\ell{\cal{D}}_\ell] 
\qquad [t_k,{\cal{D}}_\ell] = 0 \\
&=& c_kc_\ell\left([t_k,t_\ell]{\cal{D}}_k{\cal{D}}_\ell +
t_\ell t_k[{\cal{D}}_k,{\cal{D}}_\ell]\right)
\end{eqnarray}
where
\begin{subequations}
\begin{eqnarray}
& &[t_k,t_\ell] = c_{k\ell}^{~~s}t_s \quad~~ 
t_kt_\ell = \gamma_{k\ell}^{~~s}t_s + \gamma_{k\ell} ~~~
\gamma_{k\ell} = \frac{1}{2} \delta(k,\pm)\delta(\ell,\mp)  \\
& &[{\cal{D}}_k,{\cal{D}}_\ell] = \hat c_{k\ell}^{~~s}{\cal{D}}_s
\quad {\cal{D}}_k{\cal{D}}_\ell = \hat\gamma_{k\ell}^{~~s}
{\cal{D}}_s 
\end{eqnarray}
\end{subequations}
\begin{eqnarray}
[\tau_k,\tau_\ell] &=& \frac{c_kc_\ell}{c_s}~\hat C_{k\ell}^{~~s}
\tau_s +
c_kc_\ell\gamma_{\ell k}\hat c_{k\ell}^{~~s}{\cal{D}}_s \\
\hat C_{k\ell}^{~~s} &=& c_{k\ell}^{~~s}\hat\gamma_{k\ell}^{~~s} +
\gamma_{\ell k}^{~~s}\hat c_{k\ell}^{~~s}
\end{eqnarray}
The coefficients $c_{k\ell}^{~~s}$ and $\gamma_{k\ell}^{~~s}$ are
numerically valued while $\hat c_{k\ell}^{~~s}$ and
$\hat\gamma_{k\ell}^{~~s}$ are functions of $\bar bb$.  The
structure coefficients $(\hat C,\hat c)$ 
of the $\tau$-algebra are therefore
functions of $\bar bb$.
\vskip.3cm

\no (b) {\bf Gauge Fields and Field Strengths.}

\vskip.2cm

We introduce the four vector boson fields by defining$^5$
\be
\not{\cal{W}}_{rs} \equiv ig\not\vec W\vec\tau_{rs} + ig_0
\not W^0(\tau_0)_{rs}
\ee
where
\begin{eqnarray}
(\tau_\pm)_{rs} &=& c_\pm(t_\pm)_{rs} {\cal{D}}_\pm \nonumber \\
(\tau_3)_{rs} &=& c_3(t_3)_{rs} {\cal{D}}_3 \\
(\tau_0)_{rs} &=& c_0(t_0)_{rs} {\cal{D}}_0 \nonumber
\end{eqnarray}
and $(\vec W,W^0)$ replace the components of the standard boson field
while ${\cal{W}}$ lies in the internal algebra.  Here
$({\cal{D}}_+,{\cal{D}}_-,{\cal{D}}_3,{\cal{D}}_0) \equiv
(\bar b^3\bar a^3,a^3b^3,f_3(\bar bb),1)$.

The covariant derivative is now
\be
\not\nabla_{rs} = \delta_{rs}\not\partial + \not{\cal{W}}_{rs}
\ee
The corresponding field strengths are
\be
{\cal{W}}_{\mu\lambda} = (\nabla_\mu,\nabla_\lambda)
\ee
\vskip.3cm

\no (c) {\bf Boson-Fermion Interactions.}

\vskip.2cm

We shall introduce the direct boson-fermion interactions as follows:
\be
(\bar\psi_A)_{ri}~\not\nabla_{rs}U_A(\psi_A)_{si^\prime} \quad
A = 1,2~~(r,s) = (\nu,\ell)~~\mbox{or}~~(u,d)~~\mbox{and}~~
i,i^\prime = 1,2,3
\ee
where $A=1$ labels the $(\nu,\ell)$ doublet and $A=2$ labels the
quark doublet $(u,d)$.

The $U_A$ are unitary matrices.
The form of $U_1$ is restricted by the ``universal Fermi interaction",
while $U_2$ replaces the Kobayashi-Maskawa matrix.  Here $U_2$
``rotates" the initial state $(\psi_A)_{si^\prime}$.
\be
\mbox{Here is a tentative choice:} \qquad
U_1 = 1 \qquad U_2 = e^{i\varphi(a+\bar a)}
\ee

\vskip.5cm

\section{The Field Invariant.}

We choose as the invariant of the non-Abelian vector field
\be
I = \langle 0|{\rm Tr}~{\cal{W}}_{\mu\lambda}{\cal{W}}^{\mu\lambda}|
0\rangle
\ee
which differs from the standard $I$ in the use of an expectation value
over $|0\rangle$ and the meaning of ${\cal{W}}_{\mu\lambda}$.

The non-Abelian contribution to the field strength is
\begin{subequations}
\begin{eqnarray}
{\cal{W}}_{\mu\lambda} &=& [\nabla_\mu,\nabla_\lambda] \\
&=& [\partial_\mu+{\cal{W}}_\mu ,\partial_\lambda + {\cal{W}}_\lambda]
\end{eqnarray}
where
\be
{\cal{W}}_\mu = W_\mu^s\tau_s  \qquad s = (+,-,3)
\ee
\end{subequations}
By (13.2) and the $\tau$-algebra one finds
\be
{\cal{W}}_{\mu\lambda} = W_{\mu\lambda}^{~~s}\tau_s +
\hat W_{\mu\lambda}^{~~s}{\cal{D}}_s \qquad
s = (+,-,3)
\ee
where the new field strengths are
\begin{eqnarray}
W_{\mu\lambda}^{~~s} &=& ig(\partial_\mu W_\lambda^s-\partial_\lambda
W_\mu^s) - g^2 \frac{c_mc_\ell}{c_s} \hat C_{m\ell}^{~~s}W_\mu^m 
W_\lambda^\ell \\
\hat W_{\mu\lambda}^{~~s} &=& -\frac{1}{2} g^2 c_mc_\ell\hat c_{m\ell}
^{~~s}W_\mu^m W_\lambda^\ell \delta(m,\pm)\delta(\ell,\mp) 
\delta(s,3) \\
\hat C_{m\ell}^{~~s} &=& c_{k\ell}^{~~s}\hat\gamma_{k\ell}^{~~s} +
\gamma_{\ell k}^{~~s}\hat c_{k\ell}^{~~s}
\end{eqnarray}
and
\[
I = \langle 0|{\rm Tr}[W_{\mu\lambda}^{~~s}\tau_s +
\hat W_{\mu\lambda}^{~~s}{\cal{D}}_s][W^{\mu\lambda k}\tau_k +
\hat W^{\mu\lambda k}{\cal{D}}_k\|0\rangle
\]
by (13.1) and (13.3).

Here the $W_{\mu\lambda}^{~~s}$ and $\hat W_{\mu\lambda}^{~~s}$ are
functions of $\bar bb$.  In terms of the new field strengths
$W_{\mu\lambda}^{~~s}$ and $\hat W_{\mu\lambda}^{~~s}$ the field
invariant becomes
\be
\begin{array}{rcl}
I &=& {\rm Tr}\langle 0|W_{\mu\lambda}^{~~s}W^{r\mu\lambda}
\tau_s\tau_r + \hat W_{\mu\lambda}^{~~s}\hat W^{r\mu\lambda}
{\cal{D}}_s{\cal{D}}_r + (W_{\mu\lambda}^{~~s}\tau_s)
(\hat W^{r\mu\lambda}{\cal{D}}_r) \nonumber\\  
& & \hskip4.2cm + (\hat W_{\mu\lambda}^{~~s}{\cal{D}}_s)
(W^{r\mu\lambda}\tau_r)
|0\rangle  \\
\end{array}
\ee
Since ${\rm Tr}~\tau_s=0$, $I$ may be reduced to the following
expression:
\be
I = {\rm Tr}\langle 0|W_{\mu\lambda}^{~~s}W^{\mu\lambda r}
\tau_s\tau_r + \hat W_{\mu\lambda}^{~~s}
\hat W^{\mu\lambda r}{\cal{D}}_s{\cal{D}}_r|0\rangle
\ee
Note that $W_{\mu\lambda}^{~~s}\sim\delta(s,3)$ by (13.5).
After the insertion of a complete set
of intermediate states, the field invariant becomes
\be
I = \sum_n\langle 0|W_{\mu\lambda}^{~~s}W^{\mu\lambda r}|n\rangle
\langle n|{\rm Tr}~\tau_s\tau_r|0\rangle + \sum_n
\langle 0|\hat W_{\mu\lambda}^{~~s}\hat W^{r\mu\lambda}|n\rangle
\langle n|{\rm Tr}~{\cal{D}}_s{\cal{D}}_r|0\rangle
\ee
This expression simplifies since $W_{\mu\lambda}^{~~s}$ and
$\hat W_{\mu\lambda}^{~~s}$ are functions of $\bar bb$, and
therefore have no off-diagonal elements.  Then
\be
I = \langle 0|W_{\mu\lambda}^{~~s}W^{\mu\lambda r}|0\rangle
\langle 0|{\rm Tr}~\tau_s\tau_r|0\rangle +
\langle 0|\hat W_{\mu\lambda}^{~~s}\hat W^{\mu\lambda r}|0\rangle
\langle 0|{\rm Tr}~{\cal{D}}_s{\cal{D}}_r|0\rangle
\ee
where
\begin{eqnarray}
\langle 0|{\rm Tr}~\tau_s\tau_r|0\rangle &=& c_sc_r
\langle 0|{\rm Tr}~t_st_r{\cal{D}}_s{\cal{D}}_r|0\rangle \nonumber \\
&=& c_sc_r({\rm Tr}~t_st_r)\langle 0|{\cal{D}}_s{\cal{D}}_r|0\rangle
\end{eqnarray}
$I$ is now reduced to
\be
I = \sum_{s,r}\langle 0|A_{sr}W_{\mu\lambda}^{~~s}W^{\mu\lambda r}
+ 2\hat W_{\mu\lambda}^{~~s}\hat W^{\mu\lambda r}|0\rangle
\langle 0|{\cal{D}}_s{\cal{D}}_r|0\rangle
\ee
where
\begin{eqnarray}
A_{sr} &=& c_sc_r~{\rm Tr}~t_st_r \nonumber \\
&=& c_sc_r[\delta(s,\pm)\delta(r,\mp) + 2\delta(s,3)\delta(r,3)]
\end{eqnarray}

We have now reduced (13.1) to (13.12) with the following properties.
In this expression $\langle 0|{\cal{D}}_s{\cal{D}}_r|0\rangle = 0$
unless $s$ and $r$ represent either opposite or zero charge, so
that ${\cal{D}}_s{\cal{D}}_r$ is neutral.

$W_{\mu\lambda}^{~~s}$ has the same form as in the standard theory,
but the structure coefficients $C_{m\ell}^{~~s}$ differ from those of
the Lie algebra of $SU(2)$ in that they are not numerically valued
but depend on $\bar bb$.

Since $I$ is evaluated on the state $|0\rangle$, however, all
expressions $F(\bar bb)$ become $F(|\beta|^2)$.  Therefore the
structure coefficients $C_{m\ell}^{~~s}(\bar bb)$ become
$C_{m\ell}^{~~s}(|\beta|^2)$.

The final reduced form of $\langle 0|{\rm Tr}~{\cal{W}}_{\mu\lambda}
{\cal{W}}^{\mu\lambda}|0\rangle$ will have one part
$\sim W_{\mu\lambda}^{~~s}W^{\mu\lambda r}$, essentially the same as the
standard theory but with structure coefficients depending on
$|\beta|^2$.

There is also a second part $\sim \hat W_{\mu\lambda}^{~~s}
\hat W^{\mu\lambda r}$ which is
also dependent on $q$ and $\beta$.  The sum of these two parts is
multiplied by $\langle 0|{\cal{D}}_s{\cal{D}}_r|0\rangle$, again a
function of $q$ and $\beta$, the two parameters of the theory.  These
expressions also depend on the numerical coefficients 
$(c_+,c_-,c_3,c_0)$ introduced in the definition of the $\tau$.
The dependence of these coefficients on $q$ and $\beta$ will be fixed 
in the Higgs sector.

\vskip.5cm

\section{Gauge Invariance.}

The new gauge group is generated by the following 
unitary transformations:
\be
{\cal{S}} = S\otimes s
\ee
where $S$ is the standard symmetry:
\be
S \in \mbox{local}[SU(2)\otimes U(1)]
\ee
and $s$ is the gauge symmetry of the knot:
\be
s \in \mbox{global}~ U_a(1)\otimes U_b(1)
\ee
or
\be
S = e^{i\vec t\vec\theta(x)}e^{it_0\theta_0(x)}
\ee
and
\be
s = e^{iQ_a\theta_a}e^{iQ_b\theta_b}
\ee
where $\theta_a$ and $\theta_b$ are independent of $x$.
Then
\begin{eqnarray}
{\cal{S}}\psi_1 &=& {\cal{S}}\left(
\begin{array}{c}
D_\nu \\ D_\ell 
\end{array} \right) = S \otimes s\left(
\begin{array}{c}
D_\nu \\ D_\ell
\end{array} \right) \\
&=& S\left(
\begin{array}{c}
D_\nu^\prime \\ D_\ell^\prime
\end{array} \right) \\
&=& e^{i\vec t\vec\theta(x)}e^{it_0\theta_0(x)}\left(
\begin{array}{c}
D_\nu^\prime \\ D_\ell^\prime
\end{array} \right)
\end{eqnarray}
where
\be
D_k^\prime = e^{iQ_a(k)\theta_a}e^{iQ_b(k)\theta_b}D_k \qquad
k = (\nu,\ell)
\ee

The interaction terms will transform as
\be
(\bar\psi_A)^\prime\not\nabla^\prime (U_A\psi_A)^\prime =
(\bar\psi_A\bar{\cal{S}})\not\nabla^\prime ({\cal{S}}U_A\psi_A) 
\ee
Since ${\cal{S}}$ is unitary 
\be
{} = \bar\psi_A{\cal{S}}^{-1}\not\nabla^\prime({\cal{S}}U_A\psi_A)
\ee

Then the interaction terms are invariant if
\be
\not\nabla^\prime = {\cal{S}}\not\nabla{\cal{S}}^{-1}
\ee
and since $\not\nabla = \not\partial + \not{\cal{W}}$.
\begin{eqnarray}
\not{\cal{W}}^\prime &=& {\cal{S}}\not{\cal{W}}{\cal{S}}^{-1} +
{\cal{S}}\not\partial{\cal{S}}^{-1} \nonumber \\
&=& (Ss)\not{\cal{W}}(s^{-1}S^{-1}) + S\not\partial S^{-1}
\end{eqnarray}
since $s$ is global.  
The field strengths transform as
follows:
\begin{eqnarray}
{\cal{W}}^\prime_{\mu\lambda} &=& [\nabla^\prime_\mu,\nabla^\prime_\lambda] \\
&=& {\cal{S}}[\nabla_\mu,\nabla_\lambda]{\cal{S}}^{-1} 
\end{eqnarray}
Then
\be
{\cal{W}}^\prime_{\mu\lambda} = {\cal{S}} {\cal{W}}_{\mu\lambda}{\cal{S}}^{-1}
\ee
and the non-Abelian field invariant will transform as
\be
{\rm Tr}~({\cal{W}}_{\mu\lambda}{\cal{W}}^{\mu\lambda})^\prime = 
{\rm Tr} {\cal{S}}
{\cal{W}}_{\mu\lambda}{\cal{W}}^{\mu\lambda}{\cal{S}}^{-1}
\ee
where the trace is on the $t$ matrices.  Then
\begin{eqnarray}
{\rm Tr}~{\cal{S}}({\cal{W}}_{\mu\lambda}{\cal{W}}^{\mu\lambda}){\cal{S}}^{-1} &=&
{\rm Tr}~sS({\cal{W}}_{\mu\lambda}{\cal{W}}^{\mu\lambda}) S^{-1}s^{-1} \\
&=& s\cdot{\rm Tr}~S({\cal{W}}_{\mu\lambda}{\cal{W}}^{\mu\lambda})
S^{-1}\cdot s^{-1} \\
&=& s\cdot{\rm Tr}~{\cal{W}}_{\mu\lambda}{\cal{W}}^{\mu\lambda}\cdot
s^{-1}
\end{eqnarray}

Let us next expand the trace as follows:
\begin{eqnarray}
{\rm Tr}~{\cal{W}}_{\mu\lambda}{\cal{W}}^{\mu\lambda} &=& {\rm Tr}
[W_{\mu\lambda}^{~~s}W^{\mu\lambda r}\tau_s\tau_r +
\hat W_{\mu\lambda}^{~~s}\hat W^{\mu\lambda r}{\cal{D}}_s{\cal{D}}_r]
\nonumber \\
&=& c_mc_pW_{\mu\lambda}^{~~m}W^{\mu\lambda p}
{\rm Tr}~t_mt_p{\cal{D}}_m{\cal{D}}_p  \\
& &\mbox{} + 2 \hat W_{\mu\lambda}^{~~m}
\hat W^{\mu\lambda p}{\cal{D}}_m{\cal{D}}_p \nonumber
\end{eqnarray}
Then, since $W_{\mu\lambda}^{~~m}$ and $\hat W_{\mu\lambda}^{~~m}$ are functions of $\bar bb$ only and since
\be
s\bar bbs^{-1} = \bar bb
\ee
one has
\be
\begin{array}{rcl}
s{\rm Tr}~{\cal{W}}_{\mu\lambda}{\cal{W}}^{\mu\lambda})s^{-1} &=&
c_mc_pW_{\mu\lambda}^{~~m}W^{\mu\lambda p}({\rm Tr}~t_mt_p)
(s{\cal{D}}_m{\cal{D}}_ps^{-1}) \\
& &\mbox{}+ 2\hat W_{\mu\lambda}^{~~m}\hat W^{\mu\lambda p}
(s{\cal{D}}_m{\cal{D}}_ps^{-1})
\end{array}
\ee
Here ${\rm Tr}~t_mt_p$ vanishes unless $(m,p) = (\pm,\mp)$ or
$(m,p) = (3,3)$.  Hence the first term on the right 
side of (14.23) vanishes unless
$(m,p) = (\pm,\mp)$ or $(m,p) = (3,3)$ or
\be
s{\cal{D}}_m{\cal{D}}_ps^{-1} = s{\cal{D}}_\pm{\cal{D}}_\mp s^{-1}
\qquad \mbox{or} \qquad s{\cal{D}}_3{\cal{D}}_3s^{-1}
\ee
But ${\cal{D}}_3{\cal{D}}_3$ as well as ${\cal{D}}_\pm{\cal{D}}_\mp
(=\bar{\cal{D}}_\mp{\cal{D}}_\mp)$ are neutral (zero $Q_a$ and
$Q_b$ charges) and are therefore invariant under the
$s$-transformation.

The second term on the right is invariant for the same reason since
$\hat W_{\mu\lambda}^{~~m}\hat W^{\mu\lambda p}$ 
vanishes unless $(m,p) = (3,3)$.  Therefore
\be
{\cal{S}}({\rm Tr}~{\cal{W}}_{\mu\lambda}{\cal{W}}^{\mu\lambda})
{\cal{S}}^{-1} = {\rm Tr}~{\cal{W}}_{\mu\lambda}
{\cal{W}}^{\mu\lambda}
\ee

Hence the non-Abelian field invariant (13.1) is gauge-invariant.
By (14.10) and (14.12) the fermion-boson interaction terms (12.11) are also gauge-invariant.  Finally we will maintain the
gauge-invariance of the Higgs sector in the usual way.

\vskip.5cm

\section{The Higgs Sector.}

\no (a) {\bf The Vector Masses.}

The neutral coupling in the knot model may be chosen as
\be
ig\not W_3\tau_3 + ig_0\not W_0\tau_0
\ee
Introducing the physical fields ($A$ and $Z$) in the standard way we
have
\begin{eqnarray}
W_0 &=& A\cos\theta - Z\cos\theta \\
W_3 &=& A\sin\theta + Z\cos\theta
\end{eqnarray}
Then the neutral couplings (15.1) becomes
\be
{\cal{A}}A + {\cal{Z}}Z
\ee
where
\begin{eqnarray}
{\cal{A}} &=& i(g\tau_3\sin\theta + g_0\tau_0\cos\theta) \\
{\cal{Z}} &=& i(g\tau_3\cos\theta - g_0\tau_0\sin\theta)
\end{eqnarray}

Since there is no interaction between photons and neutrinos, one has
by (15.4)
\be
\left(\bar D_\nu^{3/2}~0\right) {\cal{A}} \left(
\begin{array}{c}
D_\nu^{3/2} \\ 0 
\end{array} \right) = 
\bar D_\nu^{3/2}{\cal{A}}_{\frac{1}{2}\frac{1}{2}}D_\nu^{3/2}
= 0
\ee
According to (15.5) the preceding equation is satisfied by
\be
{\cal{A}} = i(g\tau_3\sin\theta + g_0\tau_0\cos\theta) = 0
\ee
and by (15.6) and (15.8)
\be
{\cal{Z}} = ig\frac{1}{\cos\theta}~\tau_3
\ee
Then the covariant derivative of any neutral state is
\be
\nabla = \partial + ig\left[W_+\tau_+ + W_-\tau_- +
\frac{Z\tau_3}{\cos\theta}\right]
\ee
Denote the neutral Higgs scalar by
\be
\varphi = \rho(x)D_\nu|0\rangle
\ee
where $D_\nu$ is the neutral trefoil; namely,
(-3,2), carrying the representation $D^{3/2}_{-\frac{3}{2}
\frac{3}{2}}$.

We now replace the kinetic energy term of the neutral Higgs of the
standard model by the corresponding term of the knot theory as
follows:
\begin{eqnarray}
& &\frac{1}{2}{\rm Tr}(\overline{\nabla_\mu\varphi}\nabla^\mu
\varphi) \\
& &\mbox{}=\frac{1}{2}{\rm Tr}\langle 0|\bar D_\nu\left[\partial_\mu\rho
\partial^\mu\rho + g^2\rho^2(W_+^\mu W_{+\mu}\bar\tau_+\tau_+ 
\right.\nonumber\\
& & \mbox{} \hskip2.8cm \left.+ W_-^\mu W_{-\mu}\bar\tau_-\tau_- 
+ \frac{Z^\mu Z_\mu}{\cos^2\theta}\bar\tau_3\tau_3)\right]
D_\nu|0\rangle \\
& &\mbox{}= I\partial_\mu\rho\partial^\mu\rho \nonumber \\
& & \mbox{}~~+ g^2\rho^2
\left(I_{++}W_+^\mu W_{+\mu} + I_{--}W_-^\mu W_{-\mu}  
+ \frac{I_{33}}{\cos^2\theta}Z^\mu Z_\mu\right)
\end{eqnarray}
where
\begin{eqnarray}
I~~ &=& \frac{1}{2}~{\rm Tr}\langle 0|\bar D_\nu D_\nu|0\rangle \\
I_{++} &=& \frac{1}{2}~{\rm Tr}\langle 0|
\bar D_\nu\bar\tau_+\tau_+D_\nu|0\rangle \\
I_{--} &=& \frac{1}{2}~{\rm Tr}\langle 0|\bar D_\nu\bar\tau_-\tau_-
D_\nu|0\rangle \\
I_{33} &=& \frac{1}{2}~{\rm Tr}\langle 0|\bar D_\nu\bar\tau_3\tau_3
D_\nu|0\rangle
\end{eqnarray}

To agree with the mass relations according to the standard theory,
the expressions (5.14)-(5.18) must be reduced to the following
\be
\partial_\mu\bar\rho\partial^\mu\bar\rho + g^2\bar\rho^2\left[
W_+^\mu W_{+\mu} + W_-^\mu W_{-\mu} + \frac{1}{\cos^2\theta}
Z^\mu Z_\mu\right]
\ee
where
\be
\bar\rho = I^{1/2}\rho
\ee
To achieve this reduction we impose the following relations
\be
\frac{I_{kk}}{I} = 1 \qquad \qquad k = (+,-,3)
\ee
or
\be
\frac{{\rm Tr}\langle 0|\bar D_\nu(\bar\tau_k\tau_k)D_\nu|0\rangle}
{{\rm Tr}\langle 0|\bar D_\nu D_\nu|0\rangle} = 1
\ee
since $\tau_k = c_kt_k\tau_k$, one has
\be
|c_k|^{-2} = \frac{\langle 0|\bar D_\nu(\bar{\cal{D}}_k{\cal{D}}_k)
D_\nu|0\rangle}
{\langle 0|D_\nu D_\nu|0\rangle} \qquad k = (+,-,3)
\ee
The coefficients $(c_\pm,c_3)$ were introduced as arbitrary functions.
They are now fixed by (15.23) as definite functions of $q$ and
$\beta$ as follows:
\begin{eqnarray}
|c_-|^{-2} &=& \frac{1}{2} |\beta|^6 \prod^3_1(1-q_1^{2t}|\beta|^2) \\
|c_+|^{-2} &=& \frac{1}{2}|\beta|^6\prod^2_0 (1-q^{2t}|\beta|^2) \\
|c_3|^{-2} &=& \langle 0|\bar D^3_{00}D^3_{00}|0\rangle
\end{eqnarray}
$c_0$ determined by (15.8) is
\be
c_0 = \frac{\langle 0|{\cal{D}}^3_{00}|0\rangle}
{\langle 0|{\cal{D}}^0_{00}|0\rangle}
\ee
where the Weinberg relation
\[
\tan\theta = \frac{g_0}{g}
\]
has been assumed.
Here ${\cal{D}}_{00}^3$ is a polynomial in
$\bar bb$, and $\langle 0|D^3_{00}|0\rangle$ is a polynomial in
$|\beta|^2$.

\vskip.3cm

\no (b) {\bf The Fermion Masses.}

The mass term of the Weinberg-Salam Lagrangian is
\be
{\cal{M}} \sim \bar L\varphi R + \bar R\bar\varphi L
\ee
One way of implementing (15.28) is to assume, in addition to the usual
$SU(2)$ assignments that $L$ is a $SU_q(2)$ trefoil and $R$ is a $SU_q(2)$
singlet while $\varphi$ is also a $SU_q(2)$ trefoil identical to $L$.

One would then find for the mass of the $n^{\rm th}$ fermion as
an excited state of the $(w,r)$ soliton the following:
\be
{\cal{M}}_n(w,r) = \rho(w,r)\langle n|
\bar D^{3/2}_{\frac{w}{2}\frac{r+1}{2}} D^{3/2}_{\frac{w}{2}
\frac{r+1}{2}}|n\rangle
\ee
The interpretation of (15.28) that leads to (15.29) is based on the
existence of 4 Higgs doublets, one for each trefoil, and implies that
the Higgs potential has 4 minima.
In the present model one may suppose that the Higgs potential is a scalar
function defined over the $SU_q(2)$ algebra and that it may be realized
as the expectation value $\langle 0|V(\bar\varphi\varphi)|0\rangle$ where
the minima of the expectation value lie at the 4 points where
$\varphi$ is a monomial labelling a trefoil.  These four trefoils may be
realized both as four fermionic solitons and also as four Higgs scalar
solitons. Instead of the unitarity gauge, where the standard 
Higgs doublet becomes $\left(\begin{array}{c} 0 \\ \rho \end{array}
\right)$, one postulates a privileged gauge where the 
four independent doublets $\Phi_i~(i=\nu,\ell,u,d)$ are
\be
\left(
\begin{array}{c}
\varphi_\nu~{\cal{D}}_\nu \\ 0
\end{array} \right),~\left(
\begin{array}{c}
0 \\ \varphi_\ell~{\cal{D}}_{\ell}
\end{array} \right),~\left(
\begin{array}{c}
\varphi_u~{\cal{D}}_u \\ 0
\end{array} \right),~\left(
\begin{array}{c}
0 \\ \varphi_d~{\cal{D}}_d
\end{array} \right)
\ee
representing the four Higgs partners of the four fermionic  knots.  Then adapting (15.28) one has
\be
{\cal{M}}_i = \bar L\Phi_i{\cal{R}}_i + \bar R_i
\bar\Phi_iL \qquad i = \nu,\ell,u,d
\ee
or
\be 
{\cal{M}}_i = \varphi_i(\bar L_iR_i + \bar R_iL_i)
(\bar{\cal{D}}_i{\cal{D}}_i)
\ee
leading to (15.29).

This speculative expression, a special case of $H = f(\bar bb)$,
is discussed in Refs. 3 and 4.

\vskip.5cm

\no {\bf Remarks.}

\vskip.3cm

The present model has been constructed to agree closely with the
standard model where both are well defined.  However, neither 
the standard model nor the knot model as here presented 
describes the origin of the fermionic spectrum or the origin of
the Kobayashi-Maskawa matrix.  In addition the number of Higgs
particles is also left indefinite in both approaches.  On the
other hand, as we have shown, the additional degrees of freedom
of the knot model provide a possible formal basis for filling these gaps.

\vskip.5cm

\no {\bf References.}

\begin{enumerate}
\item L. Fadeev and Antti J. Niemi,
Knots and Particles, hep-th/9610193.
\item L. H. Kauffman, Int. J. Mod. Phys. A{\bf 5}, 93 (1990).
\item R. J. Finkelstein, Int. J. Mod. Phys. A{\bf 20}, 6481
(2005).
\item A. C. Cadavid and R. J. Finkelstein, A{\bf 21}, 4264
(2006).
\item R. J. Finkelstein, A Field Theory of Knotted
Solitons, hep-th/0701124.
\end{enumerate}

\end{document}